# Consideraciones metodológicas sobre el uso del Impacto Normalizado en las convocatorias Severo Ochoa y María de Maetzu

*Methodological considerations on the use of the normalized impact in the Severo Ochoa and María de Maetzu programmes*

Daniel Torres-Salinas[a], Nicolás Robinson-García[b], Enrique Herrera-Viedma[c] y Evaristo Jiménez-Contreras[d]

a. Universidad de Granada y Universidad de Navarra, CTT, Gran Vía 48, Granada
b. Georgia Institute of Technology, School of Public Policy, Atlanta GA
c. Universidad de Granada, Departamento de Ciencias de la Computación e Inteligencia Artificial, Granada
d. Universidad de Granada, Departamento de Información y Comunicación Granada, Granada

**Resumen en español**:

*En el año de 2011 se lanzaron por primera vez las convocatorias Apoyo y acreditación de Centros de Excelencia Severo Ochoa y a Unidades de Excelencia María de Maetzu. Desde entonces estas acreditaciones se han convertido en uno de los ejes de la política científica española distribuyéndose 186 M€ y acreditando 26 centros y 16 unidades. A nivel bibliométrico uno de los criterios de evaluación más llamativos es la necesidad de que los investigadores garantes tengan un Impacto Normalizado de 1,5. En este trabajo analizamos críticamente el origen de este indicador bibliométrico normalizado en los años ochenta, las diferentes variantes que se han propuesto y las limitaciones de su uso en una convocatoria nacional. Finalmente se ofrecen una serie de recomendaciones prácticas para un uso más adecuado y preciso del indicador de Impacto Normalizado con fines evaluativos.*

**Resumen en inglés**:

*In 2011, the programme for Severo Ochoa Centers of Excellence and María de Maetzu Units of Excellence was launched for the first time. Since this programme has become one of the axes of the Spanish scientific policy. € 186 million have been distributed and 26 centers and 16 units have been accredited. One of the most relevant criteria for submission is the need for guarantor researchers to have a Normalized Impact of 1.5. In this work, we critically analyze the origin of this bibliometric indicator rooted in the 1980s, the different variants that have been proposed and the limitations its use in this programme have. Finally, we offer a series of practical recommendations for a more accurate use of normalized impact indicators for evaluative purposes.*

**Palabras Clave**: Indicadores bibliométricos, Citación Normalizada, Impacto Normalizado, Indicador Crown, Política Científica,

**Keywords**: Bibliometric Indicators, Normalized Citation, Normalized Impact, Crown Indicator, Research Policy,

## 1. Introducción

Las convocatorias *Apoyo y acreditación de Centros de Excelencia Severo Ochoa y a Unidades de Excelencia María de Maeztu* forman parte del Plan Estatal de Investigación Científica y Técnica y de Innovación. Su objetivo es financiar instituciones (centros o unidades) que realizan investigación de gran impacto y que tienen además liderazgo a nivel internacional e impacto económico/social en su entorno. La primera convocatoria Severo Ochoa (SO en adelante) tuvo lugar el año 2011 y tres años más tarde, en 2014, se complementó con la María de Maetzu (MM en adelante). La diferencia entre ambas es el tamaño de los centros a acreditar y la financiación que se obtiene. En el caso de la SO se consiguen cuatro millones repartidos en cuatro años, mientras que en la MM la financiación se reduce a dos millones. En las dos convocatorias las ayudas son renovables a la finalización del período. Hasta el momento, se han presentado un total de 247 solicitudes y el pasado 21 de Enero, concluyó el plazo de la última convocatoria.

En total, en los seis años de funcionamiento, se han acreditado 26 centros y 16 unidades (véase dataset disponible en Zenodo, https://zenodo.org/record/1161753) (Tabla 1). La financiación global del programa ha sido de 184 M€, incluyendo doce renovaciones de los años 2015 y 2016. La mayor parte recae en las acreditaciones Severo Ochoa que acumula el 82% de la financiación. En la distribución por comunidades, Cataluña cuenta con 19 acreditaciones (45%) y 84 M€ (45%), Madrid 11 acreditaciones y 52 M€, País Vasco y Comunidad Valenciana tienen 4 pero el resto de Comunidades obtienen una porción testimonial y, la mayor parte, un total de 11, no ha obtenido financiación. Por áreas temáticas, destacan las Matemáticas, Ciencias Experimentales e Ingeniería (54%) y dentro de éstas la Física. En el área de Ciencias de la Vida, el 33% de las acreditaciones son de Biomedicina. La representación global de las Ciencias Sociales es mínima con tan sólo cuatro acreditaciones, todas ellas en Economía y las Humanidades no obtienen ningún centro. Es por tanto una convocatoria con enormes sesgos en lo territorial y temático.

En lo que concierne a la documentación que se solicita el aspecto que mayor interés ha suscitado han sido los criterios de evaluación y su dependencia de los indicadores bibliométricos. Entre los requisitos de obligado cumplimiento se exigen 10 investigadores garantes, 6 en el caso de las unidades MM, que tengan un proyecto de investigación activo y superen el 1,5 de Impacto Normalizado. Es importante subrayar como, por primera vez, en una convocatoria nacional se hace uso de una medida de citación normalizada, sin embargo el indicador se ha adoptado sin un proceso de reflexión previo como muestra su aplicación a nivel micro la utilización de una variante de su cálculo con notables limitaciones y sin aval por parte de la comunidad científica. Considerando la gran carga bibliométrica de las solicitudes y la importancia que estas convocatorias están tomando como uno de los ejes de la política científica española en este trabajo analizaremos el origen y evolución del indicador de Impacto Normalizado, las diferentes formas de cálculo y sus limitaciones y, finalmente, ofreceremos una serie de recomendaciones para un uso más adecuado del mismo.

**Tabla 1. Distribución del número de solicitudes concedidas y del total financiado por comunidad autónoma en las convocatorias de centros de excelencia Severo Ochoa (MM) y María de Maetzu (MM)**

| 1.1. Número de centros y solicitudes concedidas | 2011 | | 2012 | | 2013 | | 2014 | | 2015 | | 2016 | | Total |
|---|---|---|---|---|---|---|---|---|---|---|---|---|---|
| | SO | MM | SO | MM | SO | MM | SO | MM | SO | MM | SO | MM | |
| Cataluña | 4 | | 2 | | 2 | | 1 | 4 | 2 | 2 | | 2 | 19 |
| Comunidad de Madrid | 3 | | 1 | | 1 | | | 2 | | 1 | 2 | 1 | 11 |
| Comunidad Valenciana | | | 1 | | 1 | | 1 | | | 1 | | | 4 |
| País Vaso | | | | | 1 | | | | 1 | | 1 | 1 | 4 |
| Andalucía | | | 1 | | | | | | | | | 1 | 2 |
| Galicia | | | | | | | | | | | | 1 | 1 |
| Islas Canarias | 1 | | | | | | | | | | | | 1 |
| Resto comunidades (11) | | | | | | | | | | | | | 0 |
| Total | 8 | | 5 | | 5 | | 2 | 6 | 3 | 4 | 3 | 6 | 42 |

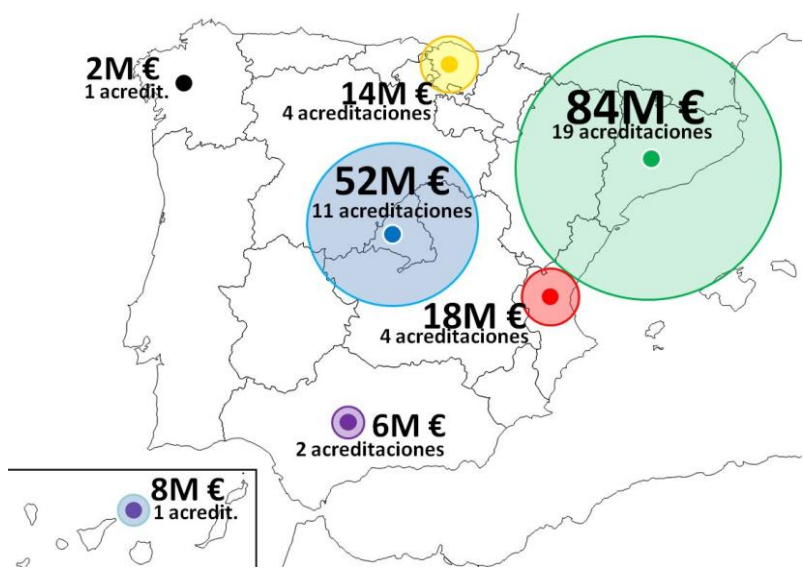

| 1.2. Total de financiación concedida (con renovaciones) | 2011 | | 2012 | | 2013 | | 2014 | | 2015 | | 2016 | | Total |
|---|---|---|---|---|---|---|---|---|---|---|---|---|---|
| | SO | MM | SO | MM | SO | MM | SO | MM | SO | MM | SO | MM | |
| Cataluña | 32M€ | | 16M€ | | 8M€ | | 4M€ | 8M€ | 8M€ | 4M€ | | 4M€ | 84M€ |
| Comunidad Madrid | 24M€ | | 8M€ | | 4M€ | | | 4M€ | | 2M€ | 8M€ | 2M€ | 52M€ |
| Comunidad Valenciana | | | 8M€ | | 4M€ | | 4M€ | | | 2M€ | | | 18M€ |
| País Vaso | | | | | 4M€ | | | | 4M€ | | 4M€ | 2M€ | 14M€ |
| Andalucía | | | 4M€ | | | | | | | | | 2M€ | 6M€ |
| Galicia | | | | | | | | | | | | 2M€ | 2M€ |
| Islas Canarias | 8M€ | | | | | | | | | | | | 8M€ |
| Resto comunidades (11) | | | | | | | | | | | | | 0M€ |
| Total | 64M€ | | 36M€ | | 20M€ | | 8M€ | 12M€ | 12M€ | 8M€ | 12M€ | 12M€ | 184M€ |

Fuente: Elaboración propia.
Acrónimos: SO - Convocatoria Severo Ochoa; MM – Convocatoria María de Maetzu

## 2. Origen y limitaciones de los indicadores normalizados

El indicador Impacto Medio Esperado empleado en la convocatoria se basa en la agregación de promedios de citas por años y disciplina de los trabajos publicados por los investigadores garantes en los últimos cuatro años comparados con la media mundial. El indicador pretende corregir sesgos en la citación causados por la edad de los documentos y las diferencias en los hábitos de citación de cada disciplina científica. Su uso, es sin duda, más correcto que otros empleados en convocatorias públicas como pueden ser los cuartiles y el Factor de Impacto (**Cabezas-Clavijo; Torres-Salinas**, 2015). Sin entrar todavía en el ámbito de aplicación se trata de un indicador ampliamente discutido en la literatura científica que parte de una larga tradición bibliométrica (**Waltman** *et al.*, 2011a). La idea de base de los indicadores de citación normalizada es obtener un valor que permita realizar comparativas entre individuos, instituciones o distintas áreas de investigación. Para ello, se hace una comparativa definiéndose dos tipos de impacto:

- Impacto Esperado: se define como el número de citas promedio que recibieron todos los trabajos del mundo publicados en el mismo año y en el mismo campo temático que el trabajo analizado.

- Impacto observado: es el número de citas que el trabajo que se trata de comparar ha recibido hasta el momento.

De la división entre el Impacto Observado y el Impacto Esperado se obtiene el Impacto Normalizado de un trabajo científico, cuyo valor será menor a 1 cuando se encuentre por debajo de la media mundial, igual a 1 cuando esté en la media mundial, o superior a 1 cuando sea mayor que la media. Uno de los principales problemas surge a la hora de agregar documentos para obtener las líneas base (baselines) sobre las que se establece la comparativa, ya que implica tener acceso completo a bases de datos bibliométricas como Web of Science o Scopus. Esto explica que sólo centros como el CWTS (Universidad de Leiden) pudieran calcular dichas líneas y el indicador a partir de los años noventa. Sin embargo las primeras tentativas para crear un indicador normalizado datan de los ochenta cuando **Schubert; Glänzel y Braun** (1983) y **Vinkler** (1986) lanzaron sus primeras propuestas. Al carecer de baselines mundiales de citación sugerían utilizar el Factor de Impacto como medida de impacto esperado por ello se considera que fueron los investigadores del CWTS (**De Bruin**, *et al.*, 1993; **Moed**, *et al.*, 1995) los primeros en sugerir sugirieron un indicador de impacto normalizado denominado CPP/FCSM (Citas por Publicación entre Promedio de Citas en el Área Temática).

El CPP/FCSM o Indicador Crown, como fue renombrado por **van Raan** (1999), se define del siguiente modo: dado un set de artículos científicos, se calcula para cada trabajo su Impacto Observado y se establece por otro lado su Impacto Esperado. El Impacto Esperado se determina considerando no sólo el año de publicación y el área temática, sino también el tipo documental (i.e., artículo, revisión o carta). Posteriormente se divide la suma total del Impacto Observado entre la suma del Impacto Esperado de todas las publicaciones (Tabla 2). Su principal limitación radica en que al ser un promedio global de citas observadas y esperadas, los campos con mayores tasas de citación pueden acaban determinando el valor final. Esta limitación motivó la principal crítica a cargo de **Lundberg** (2007) del

Karolinska Institute centraba en la necesidad de establecer los promedios entre citas observadas y esperadas para cada artículo y después sumar los totales. La Tabla 2 muestra mediante un ejemplo, las diferencias en los valores obtenidos según la formulación del indicador normalizado empleada.

**Tabla 2. Ejemplo aplicado del cálculo de la citación normalizada atendiendo a las propuestas del CWTS y del Karolinska Institute**

> Un investigador ha publicado tres trabajos:
>
> a) Determinamos tipo documental, año y categoría. Se calcula el número de citas a las que denominados Impacto Observado
> • (Trabajo A) Artículo publicado en 2014 en Comunicación con 9 citas
> • (Trabajo B) Revisión publicada en 2015 en Educación con 5 citas
> • (Trabajo C) Artículo publicado en 2016 en Comunicación con 2 citas
>
> b) Establecemos valores de referencia medios mundiales atendiendo a tipo documental, año y área a los que denominados Impacto Esperado de un trabajo
> • Los artículos de 2014 en Comunicación tienen un promedio 6 citas
> • Las revisiones de 2015 en Educación tienen un promedio de 5 citas
> • Los artículos de 2016 en Comunicación tiene un promedio de 4 citas
>
> c) Con los datos anteriores se puede calcular la citación normalizada
>
> Según la fórmula del CWTS (**Moed**, *et al.*, 1995):
> - División entre el Impacto Observado y Esperado: (9+5+2) / (6+5+4) =1,06
> - El valor es 1,06 significa que se está un 6% por encima de la media mundial.
>
> Según la fórmula del Karolinska (**Lundberg**, 2007)
> - Citas del artículo A entre su impacto esperado: 9/6= 1,5
> - Citas de la revisión B entre su impacto esperado: 5/5= 1
> - Citas de la entre su impacto esperado=2/4=0,5
> - Cálculo final del indicador: (1,5+1+0,5)/3= 1,01
> - El valor es 1,01 significa que se está un 1% por encima de la media mundial.

La crítica de **Lundberg** (2007), refrendada por **Ophtof y Leydesdorff** (2010), dio lugar a un enconado debate (**van Raan**, *et al.*, 2010) a raíz del cual los investigadores del CWTS decidieron redefinir la propuesta de Moed y desarrollar el indicador normalizado "definitivo": el MNCS o Mean Normalized Citation Score (Puntuación del Promedio de Citas Normalizada). Sin embargo lo que hicieron fue copiar la formulación de **Lundberg** (2007). La única diferencia que introducen tiene que ver con el tratamiento que hacen de las áreas temáticas ya que el MNCS aplica un factor corrector para las revistas que están en más de una categoría temática. Anteriormente una de las limitaciones del Impacto Esperado radicaba en los trabajos publicados en revistas indexadas en más de una categoría temática Web of Science introducía sesgos en los promedios de citación.

Asimismo, **Waltman** *et al.* (2011b) indica dos limitaciones adicionales del MNCS. La primera se refiere a la asimetría que caracteriza a las distribuciones de citas, que hacen que sea problemático el uso de medias aritméticas. En segundo lugar, al normalizar por años, los artículos más recientes pueden influir sustancialmente en el valor del MNCS. Posteriormente, han surgido nuevas propuestas, sobre todo enfocadas en limitar los problemas y variaciones debidas al empleo de una clasificación temática u otra. Sin embargo, la formulación de Impacto Normalizado de Waltman se ha mantenido y es la aceptada por la comunidad bibliométrica. Las variaciones en los valores obtenidos según la procedencia del indicador (*CWTS*, *InCites*, *Scival*, etc.) tienen más que ver con el trato que se hace a la multicategoría y no con los fundamentos teóricos del indicador. Por ejemplo, *InCites* calcula el promedio de impactos esperados en las distintas categorías en las que un mismo trabajo está indexado mientras que el CWTS apuesta, como se ha reseñado, por un cálculo fraccionado.

Más allá de las consideraciones técnicas del indicador y las diferencias entre las propuestas, hay una serie de características que hacen que este indicador no sea recomendable al analizar grupos reducidos de publicaciones. La principal viene dada por ser un indicador relativo que es independiente del tamaño del set de datos (se puede tener el mismo impacto normalizado con un trabajo que con 10), esto hace que se pueda premiar a investigadores muy poco productivos frente a otros más productivos que teniendo artículos de alto impacto, no son capaces de mantener los mismos niveles de citación en todos sus trabajos. De hecho, es práctica habitual, emplear este indicador junto con otros indicadores absolutos (i.e., número total de artículos publicados, número de artículos altamente citados, etc.) para facilitar su interpretación. A este tipo de limitaciones tendríamos que añadirle su poca funcionalidad en áreas de Ciencias Sociales y Humanidades dónde cualquier variación en el número citas puede provocar una subida sustancial en el indicador, al ser áreas con promedios muy bajos.

**Tabla 3. Ejemplo de cálculo del indicador de impacto normalizado según la convocatoria Severo Ochoa / María de Maeztu 2017. Periodo 2012-15 sólo se incluyen artículos y revisiones indexados en Web of Science**

| | **A** Nº Trabajos | **B** Nº Citas | **C** B/A Promedio Citas | **D1** Categoría | Baseline empleado | **D** Valor Medio | **E** C/D Comparación Valores | **F** % Trabajos Año | **G** E*F Final |
|---|---|---|---|---|---|---|---|---|---|
| 2012 | 3 | 17 | 5,67 | MEDICAL INFORMATICS | WoS | 10,61 | 0,53 | 0,18 | 0,10 |
| 2013 | 5 | 58 | 11,60 | ARTIFICIAL INTELLIGENCE | WoS | 10,09 | 1,15 | 0,31 | 0,36 |
| 2014 | 5 | 29 | 5,80 | NEUROSCIENCES | WoS | 10,23 | 0,57 | 0,35 | 0,18 |
| 2015 | 3 | 12 | 4,00 | NEUROSCIENCES | WoS | 5,91 | 0,68 | 0,18 | 0,13 |
| | | | | | | | | Impacto Normalizado→ | 0,76 |

Uno de los aspectos más destacados de las convocatorias SO y MM es que la fórmula de citación normalizada empleada (Tabla 3) no coincide con ninguna de las propuestas descritas. Para las acreditaciones SO/MM se permite el uso de *Scopus* y *Web of Science* y el propio ministerio pone a disposición los baselines de cada base de datos. Se ofrecen dos baselines por base de datos, una incluye conjuntamente artículos y revisiones y otros artículos, revisiones y actas de congresos. Aquí encontramos el primer error técnico ya que no se distingue entre las tipologías documentales y, por tanto, no se respetan las diferencias de citación entre las mismas (por ejemplo una artículo Web of Science publicado en 2013 tiene un promedio de citas de 9,54 y las revisiones tienen un promedio de 24,05). En la Tabla 4 se ofrecen algunos datos para diferentes categorías y tipologías documentales de Web of Science que evidencian la necesidad de presentar baselines desagregados por tipos documentales

*Paper published in El profesional de la información*
http://doi.org/10.3145/epi.2018.mar.15**Tabla 4. Ejemplo de las diferencias de citación que se producen en diferentes tipologías documentales y categorías Web of Science en 2013.**

| Categoría Web of Science | Promedio Citas Artículo | Promedio Citas Revisión | Promedio Citas Proceeding |
|---|---|---|---|
| Global | 9,54 | 24,05 | 0,88 |
| Neurosciences | 13,95 | 27,32 | 1,09 |
| Sociology | 3,49 | 4,14 | 0,39 |
| Ecology | 12,46 | 40,06 | 0,67 |
| Artificial Intelligence | 10,39 | 31,19 | 1,84 |

En el caso de la asignación de las categorías también se obvia las propuestas habituales que asignan a cada trabajo su categoría científica correspondiente y se normaliza en este sentido ítem a ítem. En la fórmula SO/MM el Impacto Esperado se selecciona para cada uno de los años y no para cada uno de los trabajos. Es decir la categoría donde más se publique cada año es la que se toma como referencia y en caso de empate entre categorías un año determinado se escoge la que considere el investigador más representativa, una situación que puede dar lugar a diferencias en el cálculo. Imaginemos por ejemplo que un investigador ha publicado 3 trabajos en SCIENTOMETRICS en 2013 (Tabla 5), como la revista está en dos categorías (*Information Science & Library Science* y *CS-Interdisciplinary Applications*) el investigador podría escoger cualquiera de los dos aun existiendo diferencias significativas entre las mismas. Las diferencias entre escoger una categoría u otra pueden ser sustancial, tal y como se ilustra en la Tabla 5.

**Tabla 5. Categorías de indexación de Scientometrics y ejemplo de su influencia en el cálculo del Impacto Medio Esperado de las convocatorias Severo Ochoa y María de Maetzu**

| Categorías Web of Science de la revista Scientometrics en 2013 | Artículos publicados por un investigador | Número citas | Impacto Esperado de la categoría | Cálculo final del Impacto Normalizado |
|---|---|---|---|---|
| INFORMATION SCIENCE & LIBRARY SCIENCE | 3 | 38 | 5,13 | 2,47 |
| CS, INTERDISCIPLINARY APPLICATIONS | | | 9,09 | 1,39 |

Otro aspecto problemático es dar la posibilidad de calcular el Impacto Normalizado a través de WoS o Scopus, con baselines totalmente diferentes en cuanto a la categorización y promedio de citas. Esta situación genera que ante un mismo investigador se calculen Impactos Normalizados diferentes siendo el indicador, casi siempre, superior cuando usamos Scopus. Así, se puede dar la circunstancia contradictoria de que un investigador no cumpa el criterio, y por tanto el de investigador garante, en WoS pero sí en Scopus.

Sin embargo, la casuística es mucho más amplia. La Tabla 6 muestra el impacto normalizado calculado en ambas bases de datos para una serie de investigadores de distintas áreas. Mientras que Lucas Alado presenta prácticamente el mismo valor independientemente de la fuente, José Luis Verdegay cuenta con un Impacto Normalizado en WoS de 0,91 y en Scopus de 1,51. En este caso el investigador tiene el mismo número de trabajos en WoS y Scopus en la categoría de Artificial Intelligence dónde además los promedios de citas de los baselines son similares en ambas bases de datos, sin embargo el mayor número de citas recopiladas por Scopus hace que el indicador aumente significativamente. Casos más sangrantes son los de Leandro di Stasi (más de un punto de diferencia) o Luis Fermín Capitán (casi un punto de diferencia).

**Tabla 6. Cálculo del Impacto Normalizado para diferentes investigadores de la Universidad de Granada con las bases de datos Web of Science y Scopus para el período 2012-2015**

| Investigador UGR | Área conocimiento | Impacto Normalizado Baseline WoS | Impacto Normalizado Baseline Scopus |
|---|---|---|---|
| José Luis Verdegay Galdeano | Informática | 0,91 | 1,5 |
| Jonatan Ruiz Ruiz | Ciencias de los deportes | 1,77 | 1,88 |
| Luis Fermín Capitán Vallvey | Química | 1,38 | 2,19 |
| Lucas Alados Arboledas | Ciencias de la Tierra | 1,66 | 1,67 |
| Leandro di Stasi | Psicología | 1,36 | 2,58 |

### 3. Recomendaciones

En esta nota se ha expuesto el origen y cálculo de un indicador cuyo uso se está extendiendo notablemente a nivel nacional y que, en el caso de la convocatorias SO y MM, es un requisito de imprescindible cumplimiento para los investigadores garantes limitando sustancialmente el margen de acción de los centros. Por ello se ha querido subrayar las limitaciones de la fórmula empleada a fin de que las evaluaciones sean más precisas y rigurosas. Teniendo en cuesto lo expuesto queremos ofrecer una serie de recomendaciones para un uso más acertado del indicador de citación normalizada y una evaluación más pertinente de los programas de excelencia analizados:

- Emplear indicadores bibliométricos para asesorar y evaluar solicitudes en lugar de utilizarlo como requerimiento administrativo para poder presentar una solicitud. Actualmente, tal y como está definida la convocatoria, investigadores de la altura de Kahneman, Premio Nobel en Economía, no tienen un impacto normalizado suficiente para poder ni tan siquiera presentar una solicitud.

- Apoyar el Impacto Normalizado con otros indicadores; por ejemplo aquellos que contemplan la carrera científica completa como el Índice H o reflejan mejor la excelencia científica como el número de trabajos altamente citados. De esta forma se diferenciaría entre investigadores con algunos trabajos excelentes y otros con una distribución de citas más equitativa por trabajo (**Waltman** et al. 2011b).

Asimismo debería incluirse un número mínimo de trabajos, que garantice la validez estadística, ya que ahora mismo con tres trabajos ya se podría a ser investigador garante.

- Utilizar un indicador de citación avalado por la comunidad científica, como ocurre con el *MNCS* del CWTS o el Normalized Citation Impact implementado en *InCites*, que siguen el estándar planteado por **Lundberg** (2007). De esta forma se eliminarían deficiencias manifiestas en el cálculo actual y se haría uso de un estándar internacional que goza del aval de la comunidad bibliométrica.

- Puesto que no se destinan recursos de cara a la evaluación, al contrario de lo que sucede en otros países (**Robinson-Garcia**, 2017) al menos se debería permitir que se pueda emplear el indicador ofrecido por *InCites* o *Scival*. De esta forma los gestores podrían realizar análisis diacrónicos de sus investigadores de una forma más efectiva sin depender de los baselines anuales del Ministerio. Facilitaría la verificación por parte de los evaluadores ya que, por ejemplo, desde *InCites* se pueden descargar en diferentes formatos los trabajos con su Impacto Esperado y Observado asociado. De esta forma el indicador es más fácil de calcular para el solicitante y de verificar por el evaluador. Asimismo reducirían en muchos centros los costes ya que acceso a los datos, el cálculo de indicadores personalizados y su uso conlleva un coste económico importante y un coste de conocimiento todavía mayor que parece que recae, invariablemente, en gestores de política científica y en usuarios finales, más que en los expertos.

- Considerar la singularidad de las áreas a la hora de establecer el umbral de excelencia ya que actualmente se solicita a todas superar el 1,5. Sin embargo no en todas las áreas es igual de fácil alcanzar los mínimos exigidos, por ejemplo la UE en su conjunto en el período 2012-2015 en diferentes categorías alcanza el siguiente impacto: PHYSICS (1,66), MEDICINE, GENERAL & INTERNAL (1,48), ENGINEERING - INDUSTRIAL (1,19), PHILOSOPHY (1,14). Esto significa que en física con 1,50 se supera la media mundial pero no se alcanza el estándar europeo mientras que en Filosofía con un 1,5 no solo implicaría estar por encima del mundo si no también sustancialmente por encima del estándar europeo.

- Dadas las diferencias que se producen cuando se calcula el Impacto Normalizado en función si empleamos Web of Science o Scopus sería pertinente indicar en la convocatoria para que disciplinas científicas se puede hacer uso de una fuente u otra y no dejar en manos del solicitante la decisión. No parece tener sentido que en las Ciencias Exactas, Experimentales o de la Salud se utilice una fuente como Scopus cuando, por ejemplo, toda las evaluaciones de ANECA/CNEAI en dichas áreas se articula en torno a Web of Science, produciéndose cierta contradicción y falta de alineación de las políticas de evaluación científica.

- Buscar necesariamente otro indicador u otro criterio de selección mejor adaptado a los perfiles de publicación de las Ciencias Sociales y las Humanidades así como a sus características bibliométricas. Actualmente difícilmente se podrá aprobar una unidad de en estas áreas (excepción de la economía, psicología experimental.), no solo porque no se puede aplicar a los libros sino porque al publicarse menos trabajos y al existir menos citas la variabilidad del indicador es mayor y pierde validez.

**4. Referencias**

**Cabezas-Clavijo, Álvaro; Torres-Salinas, Daniel** (2015). *Los sexenios de investigación.* Editorial UOC.

**De Bruin, R.E.; Kint, A.; Luwel, M.; Moed, H.F.** (1993). "A study of research evaluation and planning: The University of Ghent". *Research Evaluation*, v. 3, n. 1, pp. 25-41
**Lundberg, J. (2007)**. "Lifting the crown—citation z-score". *Journal of informetrics*, v. 1, n. 2, pp. 145-154.
**Moed, H.; De Bruin, R.; van Leeuwen, T.N.** (1995). "New bibliometric tools for the assessment of national research performance: Database description, overview of indicators and first applications". *Scientometrics*, v. 33, n. 3, pp. 381-422

**Opthof, Tobias; Leydesdorff, Loet** (1983). "Caveats for the journal and field normalizations in the CWTS ("Leiden") evaluations of research performance." *Journal of informetrics*, v. 4, n. 3 pp. 423-430.

**Robinson-Garcia, N.** (2017). Los resultados del REF2014 marcan el camino a seguir en la evaluación científica. *Blok de BiD*, 12 de diciembre. *http://www.ub.edu/blokdebid/es/content/los-resultados-del-ref2014-del-reino-unido-marcan-el-camino-seguir-en-la-evaluacion*

**Schubert, A.; Glänzel; W.; Braun, T.** (1983). "Relative citation rate: A new indicator for measuring the impact of publications". *Proceedings of the first national conference with international participation on scientometrics and linguistics of scientific text, Varna*. pp. 80-81.

**Van Raan, A.** (1999). "Advanced bibliometric methods for the evaluation of universities". *Scientometrics*, v. 45, n. 3, pp. 417-423.

**Van Raan, A.F.; van Leeuwen, T.N.; Visser, M.S.; van Eck, N. J.; Waltman, L.** (2010). "Rivals for the crown: Reply to Opthof and Leydesdorff". *Journal of Informetrics*, v. 4, n. 3, pp. 431-435.

**Vinkler, P.** (1986). "Evaluation of some methods for relative assessment of scientific publications". *Scientometrics*, v. 10, n. 3-4, pp. 159-177.

**Waltman, L.; van Eck, N.J.; van Leeuwen, T.N.; Visser, M.S.; van Raan, A.F.** (2011a). "Towards a new crown indicator: Some theoretical considerations". *Journal of informetrics*, v. 5, n. 1, pp. 37-47.

**Waltman, L.; van Eck, N.J.; van Leeuwen, T.N.; Visser, M.S.; van Raan, A.F.** (2011b). "Towards a new crown indicator: An empirical analysis". *Scientometrics*, v. 87, n. 3, pp. 467-481.